\documentclass[11pt]{article}
\usepackage{authblk}
\usepackage{hyperref}
\pdfoutput=1

\begin{document}

\title{Water-Oil Drainage Dynamics in Oil-Wet Random Microfluidic Porous Media Analogs}

\author[1]{Wei Xu}
\author[2]{Jeong Tae Ok}
\author[2]{Keith B. Neeves}
\author[1]{Xiaolong Yin}
\affil[1]{Petroleum Engineering, Colorado School of Mines}
\affil[2]{Chemical and Biological Engineering, Colorado School of Mines}
\maketitle

\begin{abstract}

Displacement experiments carried out in microfluidic porous media analogs show that reduced surface tension leads to a more stable displacement, which is opposite to the phenomenon observed in Hele-Shaw cells where the displacement of a more viscous fluid by a less viscous one is stabilized by surface tension. In addition, geometry of porous media is observed to play an important role. Three random microfluidic porous media analogs were made to study water-oil drainage dynamics. The geometries of our analogs were designed from Voronoi tessellations of two-dimensional space. Chip 1 is made up by randomly connected channels with a uniform width of 6 $\mu$m. Chip 2 has a Gaussian channel width distribution of 4-8 $\mu$m. Chip 3 has uniform 8 $\mu$m wide channels, with some grains removed to generate large isolated pores, or vugs. All microfluidic channels are 14 $\mu$m in depth. The fluids used are 1.5\% wt. NaCl solution dyed with 1.0\% wt. FD\&C Blue \#1 and light mineral oil. The microfluidic chips fabricated using Polydimethylsiloxane with glass covers have the internal surface treated by Trichlorosilane to achieve a uniform oil-wet condition with a contact angle of 121$^\circ{}$. The aqueous phase displaces the oil phase, with a viscosity ratio of about 1:40 and a density ratio of 1:0.85. Interfacial tension between the two fluid phases is 28.37 mN/m in fluid dynamics videos 1-5, and is reduced to 3.57 mN/m using 0.5\% wt. Ethoxylated Alcohol in fluid dynamics video 6.

Videos 1-3 show water flooding processes in Chips 1-3 with 20X magnification. Capillary number ($\mu u/\sigma$) is very small ($10^{-5}$) thus the videos are played 15 times faster than real time. It is observed that both channel size distribution (Video 2) and heterogeneity in pore size (Video 3) lead to stronger fingers and reduced displacement efficiency. Video 4 magnifies the water flooding process in Chip 3 to 100X. It is observed that the meniscus in small channels retreats as water front moves into a nearby large cavity due to the difference in the capillary forces. This action makes the non-wetting fluid front move along the path made up by chains of large pores, reducing the displacement efficiency. Contact angle hysteresis is also observed during this reciprocal action of the meniscus. Videos 5 and 6, also taken at 100X magnification, show the stabilizing effect of reduced interfacial tension. Both videos were taken in Chip 2. With the lowered interfacial tension, capillary number becomes $10^{-4}$ and the displacement becomes more uniform.

This study is supported by Research Partnership to Secure Energy for America under Grant No. 09122-29.

%The link is the fluid dynamics video:
%\href{http://ecommons.library.cornell.edu/bitstream/1813/8237/2/LIFTED_H2_EMST_FUEL.mpg}{Video}.

\end{abstract}

\end{document}